\documentstyle[multicol,pre,aps]{revtex}

\draft
\begin{document}

\title{Inattainability of Carnot efficiency in the Brownian heat engine}
\author{Tsuyoshi Hondou\footnote{e-mail: hondou@cmpt.phys.tohoku.ac.jp}}
\address{
 Department of Physics, Tohoku University \\
 Sendai 980-8578, Japan}
\author{Ken Sekimoto\footnote{e-mail: sekimoto@yukawa.kyoto-u.ac.jp}}
\address{ Yukawa Institute for Theoretical Physics, Kyoto University\\
 Kyoto 606-8502, Japan}
\onecolumn
\maketitle
\widetext
 \begin{abstract}
  We discuss the reversibility of Brownian heat engine.
  We perform asymptotic analysis of Kramers equation on  
  B\"uttiker-Landauer system and show 
 quantitatively that Carnot efficiency is inattainable even 
 in a
 fully overdamping limit.
 The inattainability is attributed to the inevitable irreversible 
 heat
 flow over the temperature boundary.
 \end{abstract}
\begin{multicols}{2}

 How efficiently can Brownian heat engine work?
 This question is important not only for the construction of
 theory of molecular motors\cite{Prost}
 but also for foundation of non-equilibrium statistical physics.
 Like Carnot cycle, Brownian heat engine can 
 extract work from the difference
 of the temperature of heat baths, where
 Brownian working material operates 
 as a transducer of thermal energy into mechanical work.
 The feature of this engine is: 1) It operates autonomously.
 2) It is driven by {\em finite} difference of the temperature of heat
 baths both of which contact with the working material simultaneously.
 Thus, 
 this engine works because the system is out of equilibrium.
 Feynman\cite{Feynman} devised what is called Feynman's ratchet 
 that can rectify
 the thermal fluctuation for work using the difference of the temperature
 of two thermal baths.
  B\"uttiker\cite{Buttiker} and
 Landauer\cite{Landauer} proposed a simpler type of
 Brownian motor and pointed out that one could extract work 
 even by the simple 
 heat engine where a Brownian particle is subject to a spatially 
 periodic 
 heat baths in a periodic potential\cite{note121}.  

 One crucial point on the Brownian engines is the 
 efficiency\cite{Feynman,Parrondo,Sekimoto,Sakaguchi,Kamegawa,HondouJ,Mag981,Shibata,ParrondoE,SekimotoJ,Matsuo,Parmeggiani,Takagi}.
 Feynman claimed that his thermal ratchet can operate reversibly,
 resulting in Carnot efficiency.
 Recently, however, some authors claimed that Feynman's claim was 
 incorrect, while
 some author supported it:
 Parrondo and Espa\~nol discussed that Feynman's ratchet should
 not work reversibly since the engine is simultaneously in
 contact with heat baths at different temperatures\cite{Parrondo}.
 Sekimoto
 devised so-called "stochastic energetics" and applied it 
 into Feynman's ratchet\cite{Sekimoto}. He showed numerically that the 
 efficiency is much less than that of Carnot. Hondou and Takagi
 showed that reversible operation of Feynman's ratchet is 
 impossible using {\em reductio ad absurdum}\cite{HondouJ}.
 Magnasco and 
 Stolovitzky studied how the engine generates motion
 with detailed analysis of its phase space\cite{Mag981}.
 On the other hand, Sakaguchi claimed that the Feynman's ratchet
 could operate reversibly by proposing a
 "stochastic boundary condition"\cite{Sakaguchi}.
 Similar result is also found in ref.\cite{Matsuo} (not on Feynman's ratchet
 but on B\"uttiker-Landauer system), on which detailed discussion 
 will be made later.
 These studies have reminded us that there is difficulty as to the
 energetic description on Brownian systems,
 because naive application of conventional energetics formulated in 
 thermodynamic and/or equilibrium system into Brownian system may 
 lead incorrect result.

 Operation of Brownian engines is done by the engines themselves
  and the engines are, therefore,
  {\em out of } equilibrium.
 To clarify the non-equilibrium nature of Brownian heat engine 
 and to find how we should apply energetics on it,
 it is important to make a quantitative analysis of the
 efficiency
  without adopting over-simplification, for analysis, that loses the function
 of a heat engine.
 Because Feynman's ratchet is somewhat complex to make a rigorous
 analysis,
 it seems suitable
 to discuss B\"uttiker-Landauer\cite{Buttiker,Landauer}
  system that is the simplest system
 of Brownian motors.
 Recently, Matsuo and Sasa analyzed the
  energetics of B\"uttiker-Landauer
 system by renormalization method\cite{Matsuo}. They claimed the system
 under quasi-static process approaches Carnot efficiency in overdamping
 limit\cite{Sakaguchi-Note}.
 Their analysis was based on a rigorous calculation starting from 
 Kramers equation, and the result is clear except 
 one point:
 They assumed that the momentum degree of freedom 
 is always 
 in equilibrium with heat bath because the system 
 is overdamping\cite{AstumianPRE}.
 This assumption is not easy for us to accept because the system is singular 
 at the transition points\cite{transition}
 where the temperature of the heat bath changes
 suddenly. 
 We conjectured that the essence of the mechanism of Brownian 
 heat engine be concentrated on this singular point and that
 the nature of these non-equilibrium engines would emerge by the analysis.
 Thus we will discuss the energetics of B\"utikker-Landauer
 system with paying our attention to the transition points.
 The result will also give us an insight about how we should apply 
 ``stochastic energetics''\cite{Sekimoto} to the overdamping 
 systems with space-dependent temperature.

 Let us consider a one-dimensional Brownian system that 
 B\"utikker and Landauer discussed,
  where working particles operate due to the broken uniformity
 of the temperature of the heat baths.
 \cite{Buttiker,Landauer}.
 While B\"utikker and Landauer started their discussion from overdamping
  equation of the system, we start from more basic standpoint of
 underdamped description, from which the overdamped equation is obtained
 by eliminating the momentum variable. 
 The probability density in phase space, $\rho(p,q)$, obeys 
 Kramers equation\cite{Kramers}:
 \[
 \frac{\partial \rho (p,q)}{\partial t}
  = - \left( \frac{\partial J_q}{\partial q} + \frac{\partial J_p}{\partial p}
  \right)
  = - K(q) \frac{\partial \rho(p,q)}{\partial p}
  - \frac{p}{m} \frac{\partial \rho(p,q)}{\partial q} 
 \]
 \begin{equation}
 +\frac{\gamma}{m}
 \frac{\partial}{\partial p}  \left[ p \rho(p,q) + m k_B T(q)
  \frac{\partial \rho(p,q)}
 {\partial p} \right],
 \label{eq:K}
 \end{equation}
 where
 $K(q) = -\frac{\partial U}{\partial q}$;
 $\gamma$, $m$, $J_q$, and $J_p$ are a friction constant, mass of a particle,
 and probability current in space and that in momentum,
 respectively\cite{gamma}.
 The potential $U(q)$ satisfies,
 $U(q) = U_r (q) +  g q$,
 where $U_r (q + L ) = U_r (q)$, $g$ $(>0)$ is a gradient of the global
  slope (load)
 and $ L $ is a period.
 The temperature has the same spatial period 
 as the potential, $T(q +L ) = T(q)$.
 In this B\"uttiker-Landauer system, there are two heat baths of which 
 the temperature is $T_h$ (for hot bath) and $T_c$ (for cold bath),
  respectively.
 Thus, there are the two transition points in a spatial period where the 
 thermal bath affecting the particle changes.
 Here we restrict ourselves to the case that 
 $\frac{T_c}{T_h} = O (1)$
 for simplicity.
 The system is known to operate as a molecular 
 engine\cite{Buttiker,Landauer,Sakaguchi,Matsuo}
 because the particle can {\em move} against global gradient 
  of the potential.
 The globally unidirectional motion was attributed to the difference
 of the temperatures of the baths, since the hot bath can activate
 working particle more than the cold bath. Suppose that two working
 particle {\em climb} the potential, where one is in a hot bath and the 
 other is in a cold bath.
 Then the working particle in a hot bath reaches the top of the potential
 hill more frequently than that in a cold bath,
 leading to the global motion in the present system.
 Thus, one can store work in proportion to the probability current.
 To make energetic analysis,
 we consider a ``replica'' particle, of which  the energy is
 $E = p^2/2m + \, U(q)$.
 Here,
 the ensemble average over the replicas corresponds 
 to thermodynamic limit\cite{Sekimoto}.

 It was shown that the efficiency of this engine can have Carnot
 efficiency if the irreversible heat transfer 
 at the transition points 
  is physically 
 negligible. Any Brownian motor is irreversible when 
 they operate with finite probability current. 
  Thus, the operation in 
 Carnot efficiency, if possible, must be in the 
 ``stalled state''  
\cite{Matsuo},
 where 
 the probability current in space disappears, $J_q (q) =0$
 (in an overdamped description) or $\int dp J_q (p,q) =0$ 
 (in an underdamped description).
  ``Quasi-static'' operation requires
 this ``stalled state''.
 Therefore, we evaluate 
 whether and how 
 the irreversible heat flows at the transition point
 by solving stationary solution of Kramers equation 
 at the stalled state(Eq.\ref{eq:K}).
 For this purpose,  we will restrict ourselves to the special region,
 $ q \in
 [-l_h, l_c]$, 
 around the transition point, $q=0$, where
  $l_h$ ($l_c$) are the width between the transition point
 and a proper point on hot (cold) bath that
 satisfies the following inequality:
 \begin{equation}
 l_{th} \ll l_x \ll L_x  \quad  (x = h \, \,
  \mbox{or} \, \,  c),
 \label{tth}
 \end{equation}
 where $L_h$ ($L_c$) is a width of hot (cold) bath 
 ($L_h + L_c = L$),
   and $l_{th}$ is a characteristic length scale of the transition region
  in which the probability density is different from that of thermal equilibrium.
 The length, $l_{th}$, 
 is the
 product of thermal velocity, $v_{th} (\sim \sqrt{k_B T/m})$,
  and the velocity relaxation time, $\tau (= m/ \gamma)$:
 $l_{th} \sim  v_{th} \tau $.
 The choice of $l_x$ does not alter the following result as long as the inequality, Eq.(\ref{tth}), is satisfied.  
 Although we will discuss only one transition region, the asymptotic 
 behavior does not differ in the other transition region.
 Hereafter we apply the normalization of the probability density $\rho(p,q)$
 as
 \cite{Note-normal} 
 \begin{equation}
  \int_{- \infty}^{\infty} dp \int_{-l_h}^{l_c} dq
  \rho(p,q) =1 .
 \label{normalization}
 \end{equation}

 Now, we will formulate the irreversible 
 heat transfer from a heat bath to the
 working particle.
  R.h.s. of Kramers equation (Eq.\ref{eq:K})
 has two parts. The first and the second terms
  are a Liouville operator on the probability density
 $\rho(p,q)$ and thus preserve 
 the energy.
 The last term is what describes the 
  energy transfer between 
 the heat bath and the particle, 
 of which the
  probability current in momentum space is written: $J_p^{irr} 
 = - \frac{\gamma}{m}
  \left[ p \rho(p,q) + m k_B T(q)
  \frac{\partial \rho(p,q)}
 {\partial p} \right]$.
 Because the probability current disappears, $J_p^{irr} =0$,
 for the probability density of equilibrium 
 $\rho(p,q) \propto \exp\{-\frac{p^2/2m + U(q)}{k_B T}\}$,
 the energy flow through $J_{p}^{irr}$  can be  sufficiently
 described 
 only where $q \in [-l_h, l_c]$.
 Average heat transfer from hot bath to the particle
  per unit time,
 $\langle \frac{d Q_h}{d t} \rangle$,
 reads
 \[
  \left\langle \frac{d Q_h}{d t} \right\rangle
 \sim \int_{- \infty}^{\infty} dp
  \int_{-l_h}^{0} dq
          \frac{ \partial E}{\partial p}{J_p^{irr}} \\
 \]
 \begin{equation}
 = - \int_{- \infty}^{\infty} dp \int_{-l_h}^{0} dq \frac{p}{m}
  \frac{\gamma}{m}
 \left[ p \rho(p,q) + m k_B T(q) \frac{\partial \rho(p,q)}{\partial p} \right].
 \label{eq:Q}
 \end{equation}
 By integration by part through momentum space, $p$, and the property
 that  $\rho(p,q)$ exponentially decreases 
 to zero 
 as $ p \rightarrow  \pm \infty$, we obtain
 \[
  \left\langle \frac{d Q_h}{d t} \right\rangle
  = - 2 \frac{\gamma}{m}  \int_{- \infty}^{\infty} dp \int_{-l_h}^{0}
  dq \left(\frac{p^2}{2 m} - \frac{k_B T(q)}{2} \right) \rho (p,q)
 \]
 \begin{equation}
  \equiv - 2 \frac{\gamma}{m} \left\langle \frac{p^2}{2m} 
    - \frac{k_B T_h}{2} \right\rangle_h \, .
 \label{total}
 \end{equation}
 This is the formula of the heat transfer from the hot bath
 to the particle
 \cite{Par2}.
 When the system 
 is in equilibrium with the heat bath, 
 the heat transfer $\langle \frac{d Q_h}{d t} \rangle$
 disappears,
 because the theory of equipartition requires
 $< p^2/2m > = k_B T /2$.
 This also shows that the energy exchange between the replica
 particle and the thermal bath is dominant only near the
 thermal transition point, $q=0$, where the average kinetic energy,
 $p^2/2m$, deviates from $k_B T/2$.

 We remark here how the energy flows around the transition
 point. As we are analyzing the stalled state,  the probability
 density $\rho(p,q)$ is stationary.
 Thus, the energy density, $\rho_E (q) = \int_{-\infty}^{\infty} dp
 (p^2/2m + U(q)) \rho (p,q), $
 is stationary.
 Because 
 there is no work in the stalled state, the conservation
 of energy requires that 
 $\frac{d <Q_h + Q_c>}{d t} =
   0$.
 It shows that the same quantity of the  
  heat from hot bath to the particle  flows
 from the particle to the cold bath:
 $ \langle \frac{d Q_h}{d t} \rangle = 
   - \langle \frac{d Q_c}{d t} \rangle $ .
 It should also be remarked that, in the stalled state,
 the efficiency vanishes except that the both sides 
 of the last equation vanishes,
 because the work, the numerator of the efficiency, 
 is absent here.
 Quasi-static operation is reversible only if the last equation vanishes.
 Therefore the  quantity 
 $ \langle \frac{d Q_h}{d t} \rangle$ sufficiently
 characterizes the operation at the stalled state  
 and thus we will analyze it in detail.
 Note that the following equality is simultaneously 
 derived
 \cite{note58},
 \begin{equation}
  \left\langle \frac{d Q_h}{dt}  \right\rangle
 =  \left. \int_{- \infty}^{\infty} dp \frac{p^2}{2m} \frac{p}{m} 
 \rho(p,q)\right|_{q=0} .
 \label{kinetictrans}
 \end{equation}
 The formula confirms that 
 the irreversible heat transfer is carried microscopically
 as a kinetic energy
 of the particle
 at a transition point.

  It is known, for example by the kinetic theory of gases\cite{Barrow}, that 
 there is finite heat transfer, $I$, in the system 
 where a Brownian particle of finite mass and friction is crossing 
 over the two regions with different temperature,
 even if the two thermal baths have no direct contact.
 It implies $I \equiv \langle \frac{d Q_h}{d t} \rangle > 0$.
  The authors
 of Ref.\cite{Matsuo} assumed that the heat transfer should disappear
 in overdamping limit, $\frac{m}{\gamma} \rightarrow 0$.
 However,  their assumption is not evident {\em a priori}.
 To reveal the validity of the assumption
 we have to perform proper energetic analysis {\em on} Kramers equation 
 which includes the degree of momentum, $p$, instead of overdamped 
 Fokker-Planck equation which lacks the degree.

 Hereafter, we will consider the asymptotic behavior of 
 the heat transfer, $I$, in the limit of overdamping process
  ($\gamma \rightarrow + \infty$ and/or $m \rightarrow 0$). 
 To find out the asymptotic behavior, it is convenient to use
  reference heat transfer $I^*$ of unit mass and friction in 
 an arbitrary set of units:
  $I^* \equiv I(m=1, \gamma=1)$\cite{note321}.
 By Eq.(\ref{total}), the reference heat transfer $I^*$ reads:
\begin{eqnarray}
 I^* & = & -2 \left\langle \frac{p^2}{2} -\frac{k_B T_h}{2} \right\rangle_h 
 \nonumber \\
    & =  &-2 \int_{-
 \infty}^{\infty} dp \int_{-l_h}^0 dq
  \left(\frac{p^2}{2} - \frac{k_B T_h}{2}\right)
  \rho^{*}(p,q) ,
 \label{eq:base}
 \end{eqnarray}
 where $\rho^*(p,q)$ is a probability density in the reference state:
 $m = 1$ and $\gamma = 1$. 
 We call a probability density $\rho$ and heat transfer of
 arbitrary mass and friction in a proper set of units 
 as the generic probability density and
 the generic heat transfer.
 Note that the following result is not altered if we have
 a different reference state. The choice of the
 values $m=1$ and $\gamma =1$ for reference state is 
 only for simplicity. 
 In the reference state, the characteristic length 
 of the transition region, $l^*_{th}$, that the probability 
 density in momentum, $p$, is out of equilibrium is:
 $l^*_{th} = v^{\ast}_{th} \tau^{\ast} =\sqrt{k_B T}$.

 To evaluate the generic heat transfer (Eq.\ref{eq:Q})
 in terms of the reference heat transfer (Eq.\ref{eq:Q}),
 we will find out the relation 
 between the generic probability density with arbitrary
 mass and friction
 $\rho(p,q)$ and the reference one $\rho^*(p,q)$.
  The potential term $K \frac{\partial \rho}{\partial p}$
 of Kramers equation, Eq.(\ref{eq:K}), is negligible
 when one discusses the asymptotic behavior
 of the overdamping\cite{Note1}.
 With stationary condition, 
 $\frac{\partial}{\partial t} =0$, on Eq.(\ref{eq:K}),
 we obtain the simple equation that describes stationary flow
 in phase space around the boundary, $q=0$:
 \begin{equation}
 \frac{p}{m} \frac{\partial \rho(p,q)}{\partial q} = \frac{\gamma}{m}
 \frac{\partial}{\partial p}  \left[ p \rho(p,q) + m k_B T(q) 
 \frac{\partial \rho(p,q)}{\partial p} \right] .
 \label{eq:K1}
 \end{equation}
 We found here that this equation has a scaling property in mass and 
 friction:
 The generic probability density $\rho(p,q)$ is expressed using the 
 probability density of the reference state $\rho^{\ast}(p,q)$.
 \begin{equation}
 \rho (p, q) = 
 c \rho^\ast \left(\frac{ p}{\sqrt{m}}, \frac{\gamma}{\sqrt{m}} q \right),
 \label{AA}
 \end{equation} 
 where the constant factor $c$ should be determined by 
 normalization (Eq. \ref{normalization})\cite{note10}.

 As $q$ departs from the transition point, $q=0$, further than the 
 characteristic length $\ell_{th}$, the probability density approaches
 that of equilibrium,
 where $\rho_h (p,q) = C_h \exp{\{- \frac{p^2}{2 m k_B T_h}}\}$ 
 (for $q \ll -l_{th}$),
 and   $\rho_c (p,q) = C_c \exp{\{- \frac{p^2}{2 m k_B T_c}}\}$
  (for $q \gg l_{th}$).
 The coefficients, $C_h$ and $C_c$, are then required to satisfy 
 the condition of the continuity of 
 probability current.
 Thus, we have
 \begin{equation}
 C_h T_h^{3/2} = C_c T_c^{3/2} ,
 \end{equation}
 which is consistent with the condition 
 derived for the overdamping limit\cite{Landauer}.
 The remaining condition that determine $C_x$ is the normalization.
 Note that the normalization of the probability density, 
 $\rho$, is satisfactorily carried out even 
 by neglecting the contribution from the transition region because
 the characteristic scale of the transition region $l_{th}$ is much smaller
 than the width $l_x$:
  $l_{th} / l_x  \ll 1$  ($x = h$ or $c$) (Eq.\ref{tth}).
 Then $C_h$ and $C_c$ are determined as:
 {\small
 \begin{equation}
 C_h =
 \frac{1}{\sqrt{2 \pi m k_B T_h}}  
  \frac{T_c}{T_c l_h + T_h l_c},
 \, \, \,  
 C_c = \frac{1}{\sqrt{2 \pi m k_B T_c}}  \frac{T_h}{T_c l_h 
 + T_h l_c}.
 \end{equation}
 }
 With these solutions and Eq.(\ref{AA}), 
 we obtain the relation between 
 the two normalized probability densities
 $\rho$ and $\rho^*$\cite{note10}:
 \begin{equation}
  \rho(p,q) = \frac{1}{\sqrt{m}} 
 \rho^* (\frac{p}{\sqrt{m}}, \frac{\gamma}{\sqrt{m}} q).
 \label{ratio}
 \end{equation}
 Note that this equation is valid even within the transition region.

 We can now express the heat transfer, $I$, 
 in terms of the reference heat transfer, $I^*$.
 We rewrite $I$ as:
 \begin{eqnarray}
  I & = & - 2 \frac{\gamma}{m} \int_{- \infty}^{\infty} dp \int_{-l_h}^{0} dq 
  \left(\frac{p^2}{2m} - \frac{k_B T_h}{2}\right)
   \rho(p,q) .
 \end{eqnarray}
 By change of variables such that $p' = \frac{p}{\sqrt{m}}$, 
 $q' = \frac{\gamma}{\sqrt{m}} q$\cite{note10}, we obtain 
 \begin{eqnarray}
 I  &  =  &
   -2  \int_{- \infty}^{\infty}  dp 
   \int_{-\frac{\gamma \,  l_{h}}{\sqrt{m}}}^{0} dq \left(\frac{p^2}{2} 
    - \frac{k_B T_h}{2}\right)
    \rho \left(\sqrt{m} p, \frac{\sqrt{m} q}{\gamma} \right). 
 \end{eqnarray}
 This yields using  Eq.(\ref{ratio}), 
 \begin{eqnarray}
 I =   -2  \int_{- \infty}^{\infty}  dp
   \int_{-\frac{\gamma \,  l_{h}}{\sqrt{m}}}^{0} dq \left(\frac{p^2}{2}
    - \frac{k_B T_h}{2}\right)
    \frac{1}{\sqrt{m}} \rho^*(p,q) .
 \label{abc}
 \end{eqnarray}
 This integrand is dominant only near the transition point,
 $q=0$, with 
 characteristic length $l^*_{th}$.
 As we are analyzing the asymptotic behavior such that
 $m \rightarrow 0$ and/or $\gamma \rightarrow \infty$, the 
  inequality, $(l^{\ast}_{th} \ll ) l_h \ll \gamma l_h / \sqrt{m} $, 
  is satisfied. Because the
  contribution from the interval  
 $q \in [-\gamma l_{h}/\sqrt{m}, -l_h]$ to the integral
 is negligible in Eq.(\ref{abc}) compared with that from  
 $q \in [-l_{h}, 0]$,
 the interval of this integral may sufficiently be
 replaced 
 by $ q \in [-l_h, 0]$.
 Using Eq.(\ref{eq:base}), we obtain one of 
 the main result of our paper\cite{note555}:
 \begin{equation}
  I \sim  - \frac{2}{\sqrt{m}} \int_{- \infty}^{\infty}  dp
   \int_{- l_{h}}^{0} dq \left(\frac{p^2}{2}
    - \frac{k_B T_h}{2}\right)  \rho^* (p, q) 
     =   \frac{1}{\sqrt{m}} I^* .
 \label{##}
 \end{equation}
 Because the characteristic length of the transition region
 vanishes in the overdamping limit, the scaling 
 property is exact asymptotically.

  From this result, we can learn that the irreversible 
 heat transfer at the transition point does not
 decrease when one takes overdamping limit,
 which is contrast to the claim in Ref.\cite{Matsuo}:
 One way to take this limit is to increase the friction constant,
 $\gamma$: Then, the heat transfer does not decrease, because
 the heat transfer $I$ does not depend on $\gamma$.
 The other way is to decrease the mass, $m$: Then, the heat 
 transfer does not decrease neither, moreover the heat transfer
 increases in the power of $1/\sqrt{m}$. The result justifies
 the intuitive estimation by  Der\'enyi and Astumian\cite{AstumianPRE}.
 The heat flow is a result of broken symmetry of probability density
 in momentum at the transition point, because the heat transfer disappears if
 the probability density is symmetric in phase space, as found by
 Eq.(\ref{kinetictrans}).
 Since an overdamped equation has no degree of freedom to describe
 the irreversible flow caused by
 the discontinuity of the temperature,
 the previous literature reached Carnot efficiency\cite{Matsuo}.

 Up to now, we have discussed how heat transfer 
 between the two heat baths behaves
 in the overdamping process. We found that the irreversible heat transfer
 does not decrease in the process.
 One finds, however, that the possible work out
  of the system may also vary
 according to the overdamping limit, because the probability current
 may vary due to the change of the parameters, $\gamma$ and $m$.
 Thus, it is not yet obvious whether non-vanishing 
 heat transfer, $I$, itself reveals
 that the system cannot attain Carnot efficiency
  in any condition including non-stalled state.
 Thus, in addition to the irreversible heat transfer discussed above,
  we will estimate work and work-induced heat transfer 
 in an overdamping process. 

 We will return to the original Kramers equation (Eq.\ref{eq:K}) for 
 B\"uttiker-Landauer system.
 We have analyzed this equation with retaining 
 both degrees of freedom, $p$ and $q$.
 However, we do not have to consider a momentum degree
 of freedom when we  discuss
  work out of the Brownian system,
 because the work is the function only of the 
 displacement of the position.
  Thus, we start the evaluation of the work by 
 the overdamped  Fokker-Planck equation
 for the probability density $P(q)$ of the
 system:
 \begin{equation}
  \frac{\partial P(q)}{ \partial t} = - \frac{\partial }{\partial q}  J(q)
  = \frac{1}{\gamma}  \frac{\partial}{\partial q} 
 \left[ \frac{\partial U(q)}{\partial q} 
       + \frac{\partial ( k_B T(q)) }{\partial q} \right] P(q) ,
 \label{FP}
 \end{equation}
 where the periodic boundary condition is applied: $P(0) =P(L)$ and  
 $ \frac{d P}{dq}|_{q=0} = \frac{d P}{dq}|_{q=L} $.
 Explicit mass dependence on the displacement of
  the system disappears in the overdamping limit.
 In stationary state,
  $ \frac{\partial P(q)}{ \partial t} =0 $,
  the probability current,  $J(q)$, is independent of $q$.
 The probability current $J$ reads
 \begin{equation}
  J = - \frac{1}{\gamma} \left[ \frac{\partial U(q)}{\partial q} 
 + \frac{\partial (k_B T(q))}{ \partial q}  \right] P(q) .
 \label{eq:J}
 \end{equation}
 The equation for $P(q)$ reads
 \begin{equation}
 \frac{\partial}{\partial q} \left[  \frac{\partial U(q)}{\partial q}
       + \frac{\partial (k_B T(q))}{\partial q} \right] P(q) = 0 .
 \end{equation}
 This equation shows that the change of the friction constant, $\gamma$,
 does not alter the probability density, $P(q)$.
  Thus, with Eq.(\ref{eq:J}), the probability current, $J$,
 scales as:  
 $J \propto \gamma^{-1}.$ 
 For a fixed
 load potential, the work per unit time
 $\frac{dW}{dt}$ is proportional to its probability current.
 Thus the sole operation, $\gamma \rightarrow \infty$,
 does not lead the system to Carnot efficiency, because
 the induced work $(\propto J)$ decreases while
 the irreversible heat (Eq. (\ref{##})) does not decrease.

 To find out the mass dependence on the work, 
 we consider the working particle obeying the Stokes' law
 with its radius 
 $r_B$, where the mass and the friction are specified 
 by one parameter, $r_B$:
  $\gamma \propto r_B$ and $m \propto r_B^3$. 
 Thus, we have 
 $\frac{dW}{dt} \propto J \propto \gamma^{-1} \propto r_B^{-1}$.
 The irreversible heat transfer
  $\frac{d Q_{irr}}{dt}$ that is independent of work is just 
 the heat transfer, $I$ (Eq.(\ref{##})).
 Thus we have 
 $ \frac{d Q_{irr}}{dt} \propto  m^{-1/2} \propto r_B^{-3/2}.$ 
 The work-induced heat transfer $Q_W$ that is proportional to the work 
 $W$ is proportional to the probability current $J$\cite{Matsuo}. 
 Thus, we obtain
 $\frac{d Q_W}{dt} \propto J \propto r_B^{-1}$.
 The three components determines the
 efficiency. Thus, we have 
 \begin{equation}
 \eta = \frac{\frac{d W}{dt}}{\frac{ d Q_W}{dt} + \frac{ dQ_{irr}}{dt} } 
 =  \frac{ c_1 r_B^{-1}}{c_2 r_B^{-1}
  + c_3 r_B^{-3/2}}
 = \frac{c_1}{c_2 + \frac{c_3}{\sqrt{r_B}}} ,
 \end{equation}
 where $c_1$, $c_2$ and $c_3$ are constants.
 The result tells that the efficiency decreases monotonically
 to zero  when one takes
 overdamping limit $r_B \rightarrow 0$.
 This result is not altered even if one includes another transition
 point in the same period, because
  the asymptotic behavior of the
 two are the same.

  In this paper, we have analyzed the energetics of
  a Brownian motor 
 of B\"uttiker-Landauer type.
  We showed quantitatively  that
 irreversible heat transfer does not disappear even if one takes 
 overdamping limit ($\gamma \rightarrow + \infty$ and/or
  $m \rightarrow 0$). This result is in contrast to the claims 
  by Refs.\cite{Sakaguchi,Matsuo}.
 The mass dependence on the irreversible heat is consistent to 
 the intuitive estimation by Ref.\cite{AstumianPRE}.
 We further analyzed the effect of non-vanishing irreversible heat transfer
 on the efficiency and showed that, even in fully overdamping limit,
 Carnot efficiency is 
 inattainable 
  for the particle obeying Stokes' law.
 It shows that the maximum efficiency of the Brownian
 motor is not attained in the stalled state.
 The result revealed that the Brownian heat engine is qualitatively
 different from heat engines of which the most efficient operation
 is in quasi-static:
 Quasi-static process is the worst condition for the
 Brownian heat engine to work, while is the best for 
 Carnot cycle.

 The location of irreversible heat transfer is the transition
 region characterized by the thermal length
 $l_{th}$.
 It is certain that the characteristic length $l_{th}$ may disappear
  in the fully overdamping limit. Then, however, 
 the  irreversible effect
 at the transition region cannot be eliminated. 
   From the result we also learn a lesson how to apply energetics
 to overdamping systems with a space-dependent temperature:
 We should apply energetics {\em before} taking
 overdamping limit.
 Otherwise, we might fail in proper evaluation of 
 irreversible heat transfer within 
 the transition region\cite{Sakaguchi,Matsuo},
 because energetic interaction between the heat bath and the particle
 is carried by the momentum exchange 
 between them. When the particle has  smaller kinetic energy
 than that expected by equipartition theorem, the particle
 receive kinetic energy from the heat bath in average.
 Thus, if we lose the degree of momentum as in the
 overdamped equation, we cannot describe this existing physical
 process properly.

 The present system cannot have maximum efficiency
 at a quasistatic condition.
  This means that the maximum efficiency is achieved 
 with finite probability current, which is therefore accompanied
 by irreversible dissipation. 
 Thus, 
 the next challenging question would be, 
 ``Is there any principle 
 that determines 
 the optimal efficiency 
 in Brownian heat engines?''

 We would like to thank T. Chawanya, T. Mizuguchi, H. Hayakawa, 
 H. Nishimori, M. Matsuo, S. Sasa, T. Shibata, Y. Hayakawa, M. Sano,
  F. Takagi, Y. Sawada, and T. Tsuzuki for helpful discussion.
 This work is supported in part
 by the Japanese Grant-in-Aid for Science Research Fund from the Ministry
 of Education, Science and Culture (Nos. 09740301, 11156216 and 12740226)
  and the Inamori Foundation.

 \end{multicols}

\begin{thebibliography}{99}
\bibitem{note121}
 Energetics was not discussed quantitatively at that time, 
 because little was known in the energetic evaluation
 of Brownian systems.
\bibitem{Sakaguchi-Note}
 The similar claim and assumption is found in Ref.\cite{Sakaguchi}.
\bibitem{transition}
  The term ``transition'' was introduced in Ref.\cite{Landauer} in the 
  sense that the probability density in overdamping limit changes 
  suddenly.
\bibitem{gamma}
  In this paper, we discussed the case that friction coefficient $\gamma$
 is independent of the temperature $T$ for simplicity.
  Friction coefficient  may depend on the
 temperature as discussed in Ref.\cite{Landauer}.
 However, it does not alter the central result, because asymptotic
 property in overdamping process is the same
 for temperature dependent friction coefficient.
\bibitem{Note-normal}
  The choice of the interval, $q \in [-l_h, l_c]$, changes the
    factor of the probability density $\rho(p, q)$. However,
       the choice does not alter the following asymptotic estimation
	 in an overdamping process.
\bibitem{Par2}

	   This formula is consistent with that derived in a different way in
	    Ref. \cite{Parrondo}.
\bibitem{note321}
    We require the reference state to hold Eq.(\ref{tth}):
     $l^{\ast}_{th} \ll l_h \ll L_h$  .
\bibitem{note10}
 Here, the parameters, $m$ and $\gamma$, are the ratios of those 
 between in the generic state
 and in the reference state. Thus, they are 
 dimensionless.
\bibitem{note555}
 The same result can also be obtained using Eq.(\ref{kinetictrans})
 instead of Eq.(\ref{total}).
\end{thebibliography}
 \end{document}